\documentclass[twoside,a4paper]{iopart}
\usepackage{epsfig}
\usepackage{amssymb}

\begin{document}

\title[Scale Dependent Metric and Minimal Length in QEG]{Scale 
Dependent Metric and Minimal Length\\ in QEG
\footnote{Talk given by M.R. at IRGAC 2006, Barcelona, Spain, July 11-15, 2006.}} 
\author{Martin Reuter and Jan-Markus Schwindt}
\address{ 
Institute of Physics, University of Mainz, D-55128 Mainz, Germany}

\begin{abstract}
The possibility of a minimal physical length in quantum gravity is discussed
within the asymptotic safety approach. Using a specific mathematical model
for length measurements (``COM microscope'') it is shown that the spacetimes
of Quantum Einstein Gravity (QEG) based upon a special class of renormalization
group trajectories are ``fuzzy'' in the sense that there is a minimal coordinate
separation below which two points cannot be resolved.
\end{abstract}

\section{Introduction}

It is an old speculation \cite{garay}
that quantum gravity induces a lower bound on physically realized distances.
Since this issue can be addressed only in a fundamental quantum theory of gravity
(as opposed to a low energy effective theory) it is natural to analyze it within 
Quantum Einstein Gravity (QEG). This theory is an attempt at the nonperturbative 
construction of a predictive quantum field theory of the metric by means 
of a non-Gaussian renormalization group (RG) fixed point \cite{wein}-\cite{max}.
From what is known today it appears indeed increasingly likely that there 
does exist
an appropriate fixed point which makes QEG nonperturbatively renormalizable
or ``asymptotically safe" \cite{souma,oliver1,oliver2,oliver3}.
The asymptotic safety scenario for QEG is most conveniently formulated
in the language of Wilson's general framework  of renormalization
\cite{wil}, using an ``exact renormalization group equation'' which defines
an RG flow on the infinite dimensional ``theory space'' consisting of all
action functionals satisfying certain symmetry constraints. The key idea is to 
base the construction of the theory on a trajectory running inside the unstable
manifold (ultraviolet critical hypersurface) of a non-Gaussian fixed point 
of the RG flow. In the extreme ultraviolet (for the RG scale $k \rightarrow
\infty$) it starts infinitely close to the fixed point, and by successive coarse
graining steps it is driven away from it, thus lowering the scale $k$. 
Conversely, starting at some finite $k$ and increasing the energy or momentum 
scale, the trajectory gets attracted into the fixed point. As a result
of this ``benign'' high energy, i.e. short distance behavior the theory is
asymptotically (i.e. for $k \rightarrow\infty$) safe from unphysical divergences
\cite{liv}.
 
An important tool in analyzing the RG flow of QEG is the effective average action
and its exact functional RG equation \cite{avact,ym}. In the case of QEG \cite{mr},
the average action is a diffeomorphism invariant functional of the metric,
$\Gamma _k [g_{\mu\nu}]$, which depends on a variable infrared (IR) cutoff $k$.
For $k \rightarrow\infty $ it approaches the bare action $S$, while it equals the 
standard effective action at $k=0$.
At least in Euclidean non-gauge 
theories on flat space,  
$\Gamma _k$ at intermediate scales has the following
properties \cite{avactrev}: (i) It defines an effective field theory at the 
momentum scale $k$. This means that every physical process which involves only a 
single momentum scale, say $p$, is well described by a tree level evaluation of
$\Gamma _k$ with $k$ chosen as $k=p$. (ii) At least heuristically \cite{avactrev},
$\Gamma _k$ may be interpreted as arising from a continuum version of a
Kadanoff-Wilson block spin procedure, i.e. it defines the dynamics of ``coarse
grained" dynamical variables which are averaged over a certain region of
Euclidean spacetime. Denoting the typical linear extension of the averaging region 
by $\ell$, one has $\ell \approx \pi /k$ in flat spacetime.
In this sense, $\Gamma _k$ can be thought of as a ``microscope" with an
adjustable resolving power $\ell =\ell(k)$.

In quantum gravity where the metric is dynamical 
the relationship between the IR
cutoff $k$ and the ``averaging scale" $\ell$ is more complicated in general. 
In the following we shall review a concrete definition of 
an ``averaging" or ``coarse graining" proper length scale $\ell=\ell(k)$.
Using this definition, along with certain qualitative 
properties of the RG trajectories of QEG, we shall demonstrate that the theory 
generates a minimal length scale in a dynamical way. The interpretation of this
scale is rather subtle, however. One has to carefully distinguish different 
physical questions one could ask, because depending on the question a minimal length
will, or will not become visible. Our presentation follows \cite{jan1}.

The running action $\Gamma_k[g_{\mu\nu}]$ can be obtained from an
exact functional RG equation \cite{mr}. In practice it is usually solved on a
truncated theory space. In the Einstein-Hilbert truncation, for instance, 
$\Gamma_k$ is approximated by a functional of the form
\begin{eqnarray}
\label{3in2}
\Gamma_k[g]=\left(16\pi G(k)\right)^{-1}\int d^4x\,\sqrt{g}\left\{
-R(g)+2 \Lambda (k)\right\}
\end{eqnarray}
involving a running Newton constant $G(k)$ and cosmological constant 
$\Lambda (k)$.

\begin{figure}
\centerline{\psfig{figure=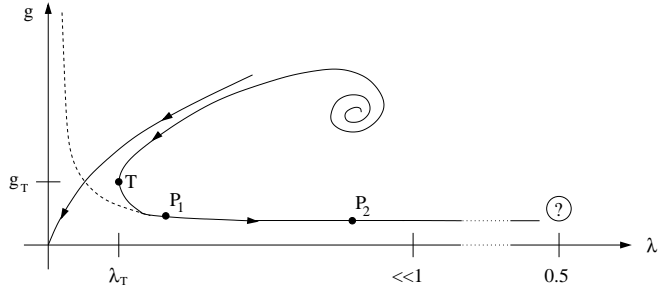, width=3.4in}}
\caption{A Type IIIa trajectory and the separatrix. The dashed line is a
trajectory of the canonical RG flow. The arrows point in the direction of 
decreasing $k$.}
\label{traj}
\end{figure}

The qualitative properties of the trajectories following from the Einstein-Hilbert
approximation are well-known by now \cite{frank1}.
Fig.~1 shows a ``Type IIIa" trajectory which would be the type that is 
presumably realized in the real universe since it is the only type that has a 
positive Newton's constant $G(k)$ and a small positive cosmological constant 
$\Lambda(k)$ at macroscopic scales. In Fig.~1 it is plotted in terms of the
dimensionless parameters 
$g(k) \equiv k^2 G(k)$ and $\lambda (k) \equiv \Lambda (k) /k^2$ 
and compared to the canonical
trajectory (dashed curve) with $\Lambda =$const and $G=$const.
The Type IIIa trajectory contains the following four parts, with
increasing values of the cutoff $k$:  \\ 
i) The classical regime for small $k$ where the trajectory 
is identical to the canonical one.
(In Fig.~1 the segment between the points $P_1$ and $P_2$.)\\
ii) The turnover regime where the trajectory, close to the Gaussian
fixed point at $g=\lambda =0$, begins to depart from the canonical
one and turns over to the ``separatrix'' which connects the Gaussian 
with the non-Gaussian fixed point $(g_*,\lambda _*)$. By definition, the 
coordinates of the turning point $T$ are $g_T$ and $\lambda _T$, and it is passed 
at the scale $k=k_T$.\\
iii) The growing $\Lambda$ regime where $G(k)$ is approximately constant but 
$\Lambda(k)$ runs proportional to $k^4$. \\
iv) The fixed point regime where the trajectory approaches the non-Gaussian fixed
point in an oscillating manner. Directly at the fixed point one has $g(k)\equiv g_*$
and $\lambda(k)\equiv\lambda _*$, and therefore $G(k) \propto k^{-2}$ and
$\Lambda (k) \propto k^2$ for $k \rightarrow\infty$. The non-Gaussian fixed point
is responsible for the nonperturbative renormalizability of the theory.\\ 
The behavior of the trajectory in the extreme infrared is not yet known since the
Einstein-Hilbert approximation breaks down when $\lambda(k)$ approaches the value
1/2. A more general truncation is needed to approximate the RG trajectory
in that region. For this reason the classical region i) does not necessarily
extend to $k=0$, and we speak about ``laboratory" scales for values of 
$k \equiv k_{\rm lab}$ in the region where $G$ and $\Lambda$ are constant. The
Planck mass is then defined as
$ m_{\rm Pl}\equiv 1/\sqrt{G(k_{\rm lab})} $.
  
In the regimes i), ii) and iii) the trajectory is well approximated by linearizing
the RG flow about the Gaussian fixed point. In terms of the dimensionful 
parameters one finds that in its linear regime $G(k)=$const and \cite{h3}
\begin{equation} \label{run}
 \Lambda(k)=\Lambda _0 \left [ 1+(k/k_T)^4 \right ]
\end{equation}
where $\Lambda _0$ is a constant. The corresponding dimensionless $\lambda
=\Lambda /k^2$ runs according to
\begin{equation}\label{run'}
 \lambda (k)=\Lambda _0 \left [ (1/k)^2+ (k/k_T^2)^2 \right ]
\end{equation}
Note that this function is invariant under the ``duality transformation"
$k \mapsto k_T^2/k$:
\begin{equation}\label{du}
 \lambda(k)=\lambda(k_T^2/k).
\end{equation}
For further details and a discussion of the other types of trajectories see
\cite{frank1,h3}. The analysis in the following sections 
refers entirely to trajectories of Type IIIa.\\

\section{Mean field metric and scale dependent distances}

Let us pick a specific RG 
trajectory, a curve $k \mapsto \Gamma _k$ on theory space.
 The effective field equations implied by 
$\Gamma _k$ define a $k$-dependent expectation value of the metric, a kind of mean
field, $\left < g_{\mu\nu}\right >_k$:
\begin{equation}\label{fe}
 \frac{\delta\Gamma _k}{\delta g_{\mu\nu}(x)}\left [ \left < g 
 \right >_k \right ]=0.
\end{equation}
In the Einstein-Hilbert truncation (\ref{3in2}) these equations are
\begin{equation}\label{fieldeq}
 R_{\mu\nu}(\left < g \right > _k)=\Lambda (k) \left < g_{\mu\nu}\right > _k.
\end{equation}  
The infinitely many equations in (\ref{fe}), one at
each scale $k$, are valid simultaneously, and all the mean fields 
$\left < g_{\mu\nu} \right >_k$ refer to one and the same physical system,
a ``quantum spacetime" in the QEG sense. The mean fields $\left < g_{\mu\nu} \right >_k$
describe the metric structure in dependence on the length scale on which the
spacetime manifold is probed. An observer exploring the structure of spacetime
using a ``microscope" of resolution $\ell(k)$ will perceive the universe as a
Riemannian manifold with the metric $\left < g_{\mu\nu} \right >_k$. While
$\left < g_{\mu\nu} \right >_k$ is a smooth classical metric at every fixed $k$,
the quantum spacetime can have fractal properties because on different scales
different metrics apply. In this sense the metric structure on the quantum
spacetime is given by an infinite set $\left \{ \left < g_{\mu\nu} \right >_k
;\; 0 \leq k< \infty\right \}$ of ordinary metrics. 

Recently it has been shown \cite{oliverfrac} that in asymptotically safe theories
of gravity, at sub-Planckian distances, spacetime is indeed a fractal whose spectral
dimension \cite{avra} equals 2. It is quite remarkable that a similar dynamical 
dimensional reduction from 4 macroscopic to 2 microscopic dimensions has also been
observed in Monte Carlo simulations of causal dynamical triangulations
\cite{ajl1,ajl2,ajl34}. (See also \cite{nino}.)

In order to understand the relation between $\ell$ and the IR cutoff $k$ we must
recall the essential steps in the construction of the average action \cite{mr}.
The formal starting point is the path integral $\int {\cal D}\gamma _{\mu\nu}
\exp \left ( -S[\gamma]\right )$ over all metrics $\gamma _{\mu\nu}$, gauge fixed by
means of a background gauge fixing condition. Even without an IR cutoff,
upon introducing sources and performing the usual Legendre 
transform one is led to an effective action $\Gamma\left [ g_{\mu\nu};
\bar{g}_{\mu\nu}\right ]$ which depends on two metrics, the expectation value of 
$\gamma _{\mu\nu}$, denoted $g_{\mu\nu}$, and the non-dynamical background field
$\bar{g}_{\mu\nu}$. The functional 
$\Gamma[g_{\mu\nu}]\equiv\Gamma[g_{\mu\nu};\bar{g}_{\mu\nu}=g_{\mu\nu}]$ 
obtained by equating the two
metrics generates a set of 1PI Green's functions for the theory.

The IR cutoff is implemented by first expanding the shifted integration variable
$h_{\mu\nu}\equiv\gamma _{\mu\nu}-\bar{g}_{\mu\nu}$ 
in terms of eigenmodes of $\bar{D}^2$,
the covariant Laplacian formed with the background metric 
$\bar{g}_{\mu\nu}$, and interpreting ${\cal D}h_{\mu\nu}$ as an
integration over all expansion coefficients.
Then a suppression term is introduced which damps the
contribution of all $\bar{D}^2$-modes with eigenvalues smaller than $k^2$.
Following the usual steps \cite{avactrev,mr} this leads to the scale dependent
functional $\Gamma_k[g_{\mu\nu};\bar{g}_{\mu\nu}]$, and again the action with one
argument is obtained by equating the two metrics:
$\Gamma_k[g_{\mu\nu}]\equiv\Gamma_k[g_{\mu\nu};\bar{g}_{\mu\nu}=g_{\mu\nu}]$.
It is this action which appears in (\ref{fe}). Because of the identification
of the two metrics it is, in a sense, the eigenmodes of
$D^2$, constructed from the argument of $\Gamma_k[g]$, which are cut off at $k^2$.
Note however that neither the $g_{\mu\nu}$- 
nor the $\bar{g}_{\mu\nu}$-argument of $\Gamma _k$
has any dependence on $k$. Therefore $\gamma _{\mu\nu}$ is expanded in terms of the
eigenfunctions of a {\it fixed} operator $\bar{D}^2$. Since its eigenfunctions are
complete, we really integrate over all metrics when we lower $k$ from infinity
to zero.
Note also that a $k$-dependent mean field arises only 
at the point where we go ``on shell" with $g_{\mu\nu}
=\bar{g}_{\mu\nu}$: the solution $\left < g_{\mu\nu} \right >_k$ to eq.~(\ref{fe})
depends on $k$, simply because $\Gamma _k$ does so.

In ref.~\cite{oliverfrac} an algorithm was proposed which allows the reconstruction
of the ``averaging" scale $\ell$ from the cutoff $k$. The input data is the set of 
metrics characterizing a quantum manifold, $\left \{ \left < g_{\mu\nu}\right >_k
\right \}$. The idea is to deduce the relation $\ell = \ell(k)$ from the spectral
properties of the {\it scale dependent} Laplacian ${\bf \Delta} _k \equiv D^2 \left ( 
\left < g_{\mu\nu}\right >_k \right )$ built with the solution of the effective
field equation. More precisely, for every fixed value of $k$, one solves the
eigenvalue problem of $-{\bf \Delta} _k$ and studies in particular the properties of the
eigenfunctions whose eigenvalue is $k^2$, or nearest to $k^2$ in the case of a discrete
spectrum. We shall refer to an
eigenmode of $-{\bf \Delta} _k$ whose eigenvalue is (approximately)
the square of the cutoff $k$ as a ``cutoff mode" (COM) 
and denote the set of all COMs by {\sf COM}($k$).

If we ignore the $k$-dependence of ${\bf \Delta} _k$ for a moment (as it would be
appropriate for matter theories in flat space) the COMs are, for a sharp cutoff,
precisely the last modes integrated out when lowering the cutoff, since the 
suppression term in the path integral cuts out all $h_{\mu\nu}$-modes with
eigenvalue smaller than $k^2$.

For a non-gauge theory in flat space the coarse graining or averaging of fields 
is a well defined procedure, based upon ordinary Fourier analysis,
and one finds that in this case the length $\ell$
is essentially the wave length of the last modes integrated out, the COMs.

This observation motivates the following 
tentative {\it definition} of $\ell$ in quantum gravity.
We determine the COMs of $-{\bf \Delta} _k$, analyze how fast these eigenfunctions vary
on spacetime, and read off a typical coordinate distance $\Delta x^\mu$
characterizing the scale on which they vary. For an oscillatory COM, for example,
$\Delta x$ would correspond to an oscillation period. 
Finally we use the metric
$\left < g_{\mu\nu} \right >_k$ itself in order to convert $\Delta x^\mu$ to a
proper length. This proper length, by definition, is $\ell(k)$.
The experience with theories in flat spacetime
suggests that the COM scale $\ell$ is a plausible {\it candidate} for a physically 
sensible resolution function $\ell =\ell(k)$, but there might also be others,
depending on the experimental setup one has in mind.

In a quantum spacetime, the (geodesic, say) distance of two given points
$x$ and $y$ depends on $k$:
\begin{equation}\label{intlen}
 L_k(x,y)\equiv\int _{{\cal C}_{xy}^{(k)}}\left (\left < g_{\mu\nu}\right >_k
 dx^\mu dx^\nu \right )^{1/2}.
\end{equation}
Here ${\cal C}_{xy}^{(k)}$ denotes the 
(possibly $k$-dependent) geodesic connecting $x$ to $y$. 
The interpretation of this $k$-dependent distance is as follows. If $k$ parametrizes
the ``resolution of the microscope" with which the spacetime is observed, the
metric $\left < g_{\mu\nu}\right >_k$ and correspondingly the distance $L_k(x,y)$
pertain to a specific scale of resolution, and different observers, using 
microscopes of different $k$-values, will measure different lengths in general. 
This $k$-dependence of lengths is analogous to the ``coastline of England 
phenomenon" well known from fractal geometry \cite{mandel,avra}.\\

\section{A mimimal length on the QEG four-sphere}

The QEG four-sphere \cite{jan1} is a manifold in the QEG sense, i.e. supplied with a 
family of infinitely many metrics $\{\left < g_{\mu\nu}\right >_k|k=0,\cdots,
\infty \}$. To be specific, it is the family of maximally symmetric solutions
of (\ref{fieldeq}) with positive curvature. It exists only provided $\Lambda(k)>0$,
which is the case for all type IIIa trajectories.

We may parametrize the $S^4$ by coordinates $(\zeta,\eta,\theta,\phi)$
with ranges $0 < \zeta,\eta,\theta < \pi$ and $0 \leq \phi < 2 \pi$. The line element
$\langle ds^2 \rangle_k \equiv \langle g_{\mu\nu}\rangle_k dx^\mu dx^\nu$ 
can be written as
\begin{equation}\label{metr}
 \langle ds^2 \rangle_k =r^2(k)\left [ d \zeta ^2 +\sin ^2 \zeta 
 \left (
 d \eta ^2 + \sin ^2 \eta ( d \theta ^2 + \sin ^2 \theta d \phi ^2 )\right ) \right ],
\end{equation}   
where $r(k)$ is the $k$-dependent radius of the $S^4$ implied by (\ref{fieldeq}):
\begin{equation}\label{rsol}
 r(k)=\sqrt{3/ \Lambda (k)}.
\end{equation}

The family of metrics (\ref{metr}), (\ref{rsol}) constitutes a concrete example of a
quantum spacetime as it was discussed in ref.~\cite{oliverfrac}.
Contrary to a Brownian curve or the coastline of England, distances {\it decrease}
when we {\it increase} the cutoff $k$. The metric scales as 
$\left < g_{\mu\nu}\right >_k \propto 1/\Lambda(k)$ so that in the fixed point 
regime $\left < g_{\mu\nu}\right >_k \propto 1/k^2$ implying $L_k(x,y)\propto 1/k$
for any (geodesic) distance.
On the equator, $\zeta =\eta =\theta =\pi /2$, the 
geodesic distance (\ref{intlen}) of two points $x$ and $y$ 
with angles $\phi(x)$ and $\phi(y)$ reads
\begin{equation}\label{geodist}
 L_k(x,y)= \sqrt{3/\Lambda(k)}\; | \phi(x)-\phi(y)|
 = k^{-1}\sqrt{3/\lambda(k)}\; | \phi(x)-\phi(y)|.
\end{equation}

On the quantum $S^4$, the scalar eigenfunctions of $-{\bf \Delta} _k$  
are the spherical harmonics $ Y_{n l_1 l_2 m}(\zeta,\eta,\theta,\phi)$,
labeled by four integer quantum numbers $n$, $l_1$, $l_2$ and $m$,
where $n \geq l_1 \geq l_2 \geq |m|$. They have the eigenvalues
\begin{equation}\label{evalues}
 {\cal E}_n= n(n+3)/r^2(k),\quad n=0,1,2,3,\cdots
\end{equation} 
The eigenvalues for the vector and
tensor modes are slightly different, but for large $n$ the 
difference becomes negligible and the spectrum is to a good approximation continuous.
We will use this continuum approximation 
since we are interested in small angular distances $\Delta\phi$
anyway. 
Let us determine the associated set of cutoff modes {\sf COM}($k$), i.e. 
the eigenfunctions with $-{\bf \Delta} _k$-eigenvalue 
as close as possible to $k^2$. Inserting ${\cal E}\approx k^2$ into 
(\ref{evalues}) and using eq.~(\ref{rsol}) for $r(k)$, we find the following
equation for the $n$-quantum number of the COMs at scale $k$:
\begin{equation}\label{nofk}
 n(k) \approx\sqrt{3/\Lambda(k)}\,k =\sqrt{3/\lambda(k)}.
\end{equation} 
Obviously $n(k)$ is indeed large if $\lambda(k)\ll 1$.
The set {\sf COM}($k$)
consists of all harmonics $Y_{nl_1l_2m}$ with $n$ fixed by eq.~(\ref{nofk})
and $l_1$, $l_2$ and $m$ arbitrary.

Apart from its obvious dependence on the scale, the set {\sf COM}($k$) depends on the RG
trajectory via the function $\lambda(k)$ which determines $n(k)$. 
The function $\lambda =\lambda(k)$ is not invertible
in general and different $k$'s can lead to the same {\sf COM}($k$).
Let us look at the Type IIIa trajectory in Fig.~1 as an example. First we concentrate
on its part close to the turning point, staying away from the spiraling regime
in the UV, and the IR region where the Einstein-Hilbert truncation breaks down.
We observe then that for every scale $k<k_T$ below the turning point there exists
a corresponding scale $k^\sharp >k_T$ which has the same $\lambda$- and therefore
$n$-value. As a result, the corresponding cutoff modes are equal at the two scales:
{\sf COM}($k$)={\sf COM}($k^\sharp$).
If the turning point is sufficiently close to the Gaussian fixed point, 
and $k$ is not too far from $k_T$, we may use the linearization (\ref{run'})
for an approximate determination of $k^\sharp$. Because of the ``duality symmetry"
(\ref{du}) it is given by
$k^\sharp =k_T^2/k$. In the ``spiraling'' regime
many different $k$-values
have the same $\lambda(k)$ and {\sf COM}($k$).
 
Next we determine the degree of position dependence of the COM's
and quantify their ``resolving power". In order to convert the 
estimate for $n(k)$, eq.~(\ref{nofk}), to
an angular resolution we note that it is sufficient
to do so for one position and one direction. By the translation and rotation symmetries
of the sphere, the resolution will be the same at any other point and in any other
direction. We therefore choose to determine the angular resolution of the modes
along the equator.

Two of the ${\bf \Delta} _k$-eigenfunctions with eigenvalue $n(k)$,
namely $Y_{\pm}\equiv Y_{nnn \pm n}$, oscillate most rapidly as a function of $\phi$,
and we shall use them in order to define the angular resolution. 
Their $\phi$-dependence is $e^{\pm in \phi}$ and
the corresponding angular resolution is
\begin{equation}\label{dphi}
 \Delta\phi(k)=\pi/n(k)=\pi\sqrt{\lambda(k)/3}.
\end{equation}
As expected, the angular resolution implied by the COMs depends on the RG trajectory.
It does so only via the function $\lambda = \lambda(k)$ and, as a result,
can be of the same size for different values of $k$. 

By definition, the
COM scale $\ell$ is the {\it proper} length corresponding to $\Delta\phi(k)$ 
as computed with the metric $\left < g_{\mu\nu}\right >_k$ of eqs.~(\ref{metr}),
(\ref{rsol}). From eqs.~(\ref{geodist}) and (\ref{dphi}) we obtain
\begin{equation}
 \ell(k)=\pi/k.  
\end{equation}
So we find that, as in theories on a classical flat spacetime, the natural 
proper length scale $\ell$ of the {\sf COM}($k$)-modes is just $\pi/k$.
Thanks to the symmetry of the sphere it is neither position nor
direction dependent. 

Taking the result $\ell\propto 1/k$, 
it seems as if nothing remarkable had happened.
But the surprising effects appear in our result for the angular resolution, 
eq.~(\ref{dphi}). As we can see from the flow diagram of Fig.~1, $\lambda(k)$ takes
on a minimum value $\lambda _T$ at the turning point $T$.
In fact, as $\lambda(k)\geq\lambda _T$ for any scale
$k$, we conclude that the angular resolution $\Delta\phi (k)$ is bounded below
by the minimum angular resolution
\begin{equation}\label{dephmin}
 \Delta\phi _{\rm min}=\pi\sqrt{\lambda _T/3}.
\end{equation}
Stated differently, there 
does not exist any cutoff $k$ for which $\Delta\phi(k)$ would be smaller
than $\Delta\phi _{\rm min}$. On the other hand, angular resolutions between 
$\Delta\phi _{\rm min}$ and 
$\Delta\phi _*\equiv\pi\sqrt{\lambda_*/3}$
are realized for at least two scales $k$.

What has happened here? Coming from small $k$, we travel along the RG trajectory 
and follow its $S^4$ solutions, 
observing spacetime with a ``microscope" of variable proper resolution $\ell(k)$.
At first, in the classical regime, an increase of $k$ leads to the resolution of
finer and finer structures since $\Lambda =$const implies $\Delta\phi(k)
\propto 1/k$. 
For the canonical RG trajectory, this behavior would
continue even for $k \rightarrow\infty$. 
In quantum gravity, however, in region ii), the sphere starts to {\it shrink},
due to a growing cosmological constant $\Lambda (k)$. At the turning point scale $k_T$ 
at which $\lambda (k)$ assumes its minimum
$\lambda _T$, the shrinking becomes faster than the improvement of the resolution
($r(k)\propto k^{-2}$ in region iii)). 
Although we can resolve smaller and smaller proper distances, this is of
no use, since the sphere is shrinking so fast that a {\it smaller} proper length 
corresponds to a {\it larger} angular distance. Finally, in the fixed point regime
(at large angles although this is an ultraviolet fixed point!),
the shrinking slows down to a rate that cancels exactly the improved resolution
of the microscope ($r(k)\propto k^{-1}$) so that the angular resolution
approaches a constant value $\Delta\phi _*$ after the oscillations have been damped 
away. 

The minimum of $\Delta\phi$ at the turning point is equivalent to a maximum of the $n$
quantum number the COMs can have:
$n_{\rm max}\approx\sqrt{3/\lambda_T}$.
This result does {\it not} 
mean that in the fundamental path integral underlying the flow equation
not all quantum fluctuations are integrated out when $k$ is lowered from infinity
to $k=0$.
It should instead be thought of as reflecting properties
of the mean field $\left < g_{\mu\nu}\right >_k$. Rather than the spectrum of the
$k$-independent operator $\bar{D}^2$ 
relevant in the path integral we analyzed that of the 
explicitly $k$-dependent Laplacian $D^2 \left ( \left < g_{\mu\nu}\right >_k \right )$;
its explicit $k$-dependence is due to the scale dependence of the on-shell metric.
Our argument reveals that the 
effective spacetime with the running on-shell metric cannot support harmonic modes
of arbitrarily fine angular resolution.

This phenomenon is a purely dynamical one; the finite resolution is not built in 
at the kinematical (i.e. $\gamma _{\mu\nu}$-) level, as it would be the case,
for instance, if the fundamental theory was defined on a lattice. 
It is also 
important to stress that, if the non-Gaussian fixed point exists, the 
Green's functions $G_n(x_1,x_2,\cdots,x_n)$
can be made well defined at all non-coincident points, i.e. 
for arbitrarily small coordinate distances among the $x_i^\mu$'s. Those
Green's functions contain information even about angular scales smaller than 
$\Delta\phi _{\rm min}$, in particular they ``know" about the asymptotic safety 
of the theory which manifests itself only at scales $k \gg k_T$.

In fact, the argument leading to the finite resolution $\Delta\phi _{\rm min}$
is fairly independent of the high energy behavior of the theory. The crucial 
ingredient in the above reasoning was the occurrence of a minimum value 
for $\lambda(k)$. This minimum occurs as a direct consequence of the $k^4$-running
of $\Lambda (k)$ given in eq.~(\ref{run}).
However, this $k^4$-running occurs already in standard perturbation theory, simply 
reflecting the quartic divergences of all vacuum diagrams. From this point of
view our argument is rather robust.

The upper bound on the angular momentum like quantum number $n$ is reminiscent
of the ``fuzzy sphere" constructed in ref.~\cite{fuzzy}. While in the case of 
the fuzzy sphere the finite angular resolution is put in ``by hand", in the present case 
it emerges as a consequence of the quantum gravitational dynamics .\\

It is instructive to ask which proper length would be ascribed to
$\Delta\phi _{\rm min}$ by an observer using the macroscopic, classical
metric $\left < g_{\mu\nu}\right >_{k_{\rm lab}}$,
where $k_{\rm lab}$ is any scale in the classical regime in which $G$ and $\Lambda$
do not run (in Fig.~1 between the 
points $P_1$ and $P_2$).
We denote this proper length by $L_{\rm min}^{\rm macro}$; it obeys
$L_{\rm min}^{\rm macro}=r(k_{\rm lab})\Delta\phi _{\rm min}$.
Using eqs.~(\ref{rsol}),(\ref{dephmin}) and (\ref{run}), and assuming
$k_{\rm lab}\ll k_T$, we obtain 
\begin{equation}\label{lmami}
 L_{\rm min}^{\rm macro}
 =\frac{\pi}{k_T}\sqrt{\frac{\Lambda(k_T)}{\Lambda(k_{\rm lab})}}
 =\frac{\pi}{k_T}\left [\frac{2}{1+(k_{\rm lab}/k_T)^4}\right ]^{1/2}
 \approx \sqrt{2}\pi k_T^{-1} .
\end{equation}
Remarkably, this minimal proper length is different in general from the Planck length
which is usually thought to set the minimal length scale.   
In fact, $L_{\rm min}^{\rm macro}$ can be much larger than $\ell _{\rm Pl}
\equiv m_{\rm Pl}^{-1}$. The trajectory realized in Nature seems to be an extreme
example: It has $k_T^{-1}\approx 10^{30}\ell _{\rm Pl}
\approx 10^{-3}$ cm, and $L_{\rm min}^{\rm macro}$ is of the same order of
magnitude.
Should we therefore expect to find an $L_{\rm min}^{\rm macro}$ 
of the order of $10^{-3}$ cm in the real world?
The answer is no, most probably. See point (ii) in the discussion at the end of the 
paper.

\section{Intrinsic distance and scale doubling}

In fractal geometry and any framework involving a length scale dependent metric
one can try to define an ``intrinsic" distance of any two points $x$ and $y$ 
by adjusting the resolving power of the
``microscope" in such a way that the length scale it resolves equals 
approximately the, yet to be determined, 
intrinsic (geodesic) distance from $x$ to $y$\footnote{This 
kind of dynamical adjustment of the resolution has also been used in the 
RG improvement of black hole \cite{bh} and cosmological \cite{cosmo1}-\cite{mof}
spacetimes, see in particular ref.~\cite{h2}.}.
To be concrete, let us fix two points $x$ and $y$ and let us try to assign to them
a cutoff scale $k \equiv k(x,y)$ which satisfies
\begin{equation}\label{self}
 L_{k(x,y)}(x,y)=\ell(k(x,y)).
\end{equation}
Eq.~(\ref{self}) is a self-consistency condition for $k(x,y)$: the LHS of (\ref{self})
is the distance from $x$ to $y$ as seen by a microscope with $k=k(x,y)$, and the
RHS is precisely the resolution of this microscope. If (\ref{self}) has a unique
solution $k(x,y)$ one defines the intrinsic distance by setting
$ L_{\rm in}(x,y)\equiv L_{k(x,y)}(x,y)$.
Since $\ell(k)=\pi/k$, this distance is essentially the inverse cutoff scale:
$ L_{\rm in}(x,y)=\pi/k(x,y)$.

Let us evaluate the self-consistency condition (\ref{self}). Without loss of 
generality we may assume again that $x$ and $y$ are located on the equator of $S^4$
so that eq.~(\ref{geodist}) applies. Then, by virtue of (\ref{dphi}), 
eq.~(\ref{self}) boils down to the following implicit equation for $k(x,y)$:
\begin{equation}\label{lamk}
 \lambda(k(x,y))=\frac{3}{\pi ^2}| \phi(x)-\phi(y)|^2.
\end{equation}    

Recalling the properties of the function $\lambda(k)$ for a Type IIIa trajectory
we see that (\ref{lamk}) does {\it not} admit a unique solution for $k(x,y)$.
If $x$ and $y$ are such that $| \phi(x)-\phi(y)|<\Delta\phi _{\rm min}$ it 
possesses no solution at all, and if $| \phi(x)-\phi(y)|>\Delta\phi _{\rm min}$
it has at least two solutions. Staying away from the deep UV and IR regimes,
every solution $k(x,y)<k_T$ on the lower branch of the RG trajectory has 
a partner solution $k(x,y)^\sharp >k_T$ on its upper branch. As a result, the 
intrinsic distance of $x$ and $y$ is either undefined, or there exist at least two 
different lengths which satisfy the self-consistency condition (\ref{self}).

In the linear regime where $k^\sharp =k_T^2/k$, the two lengths $L_{\rm in}(x,y)
=\pi /k(x,y)$ and $L_{\rm in}(x,y)^\sharp =\pi / k(x,y)^\sharp$ are related by
\begin{equation}\label{ldual}
 L_{\rm in}(x,y)^\sharp =\frac{L_T^2}{L_{\rm in}(x,y)}
\end{equation}
where $L_T \equiv \pi / k_T$. 
If $L_{\rm in}(x,y)$ is large compared to the turning point length
scale $L_T$, the ``dual" scale $L_{\rm in}(x,y)^\sharp$ is small. In the extreme
case, when applied to Nature's RG trajectory, the duality (\ref{ldual}) would
even exchange the Planck- with the Hubble-regime: $L_{\rm in}(x,y)\approx H_0^{-1}$
implies $L_{\rm in}(x,y)^\sharp \approx l_{\rm Pl}$.

This ``doubling" of $k$-scales, again, is due to
the ``back bending" of the RG trajectory at the turning point $T$ which implies 
that the function $\lambda =\lambda(k)$ assumes a minimum at a finite scale $k=k_T$.
Only the trajectories of Type IIIa possess a turning point of this kind, and this
is one of the reasons 
why they are particularly interesting and we restricted our discussion to them.

\section{Discussion}
While its origin is quite clear, the physical implications of the scale doubling
and the duality symmetry are somewhat mysterious. To some extent the difficulty
of giving a precise physical meaning to them is related to the fact that one actually
should define the ``resolution of the microscope" in terms of realistic experiments
rather than the perhaps too 
strongly idealized mathematical model of a measurement based upon the
COMs. For various reasons it seems premature to assign a direct observational 
relevance to the minimal angular resolution and the scale doubling:\\ 
(i) Only the {\it coordinate} distance $\Delta\phi(k)$ assumes a
minimum, but not the corresponding {\it proper} distance computed with the 
running metric $\left < g_{\mu\nu} \right >_k$. In particular the resolution
function $\ell(k)=\pi /k$ is exactly the same as in flat space. Nevertheless,
the COM-microscope is unable to distinguish points with an angular separation
below $\Delta\phi _{\rm min}$ !\\
(ii) Our analysis applies to pure gravity. In presence of matter the ``fuzziness"
of the $S^4$ can become visible probably only at scales where the cosmological 
constant dominates the energy density.
In particular, the fuzziness might be masked by the backreaction
of a realistic measuring apparatus on the spacetime structure.\\
(iii) As for a possible physical significance of the duality symmetry it is to be
noted that the two scales which it relates, $k<k_T$ and $k^\sharp >k_T$,
have a rather different status as far as quantum fluctuations about the mean field 
metric $\left < g_{\mu\nu}\right >_k$ are concerned. The structure of the exact
RG equation is such that the fluctuations are the larger the stronger the 
renormalization effects are. As a result, the metric fluctuations about 
$\left < g_{\mu\nu}\right >_{k^\sharp}$ on the upper branch are certainly larger
than at the dual point on the lower branch of the RG trajectory. 

Clearly more work is needed in order to understand these 
rather intriguing issues better. We
hope to return to them elsewhere.

\end{document}